\documentclass[aps,prl,print,12pt,onecolumn, footinbib,superscriptaddress,floatfix]{revtex4-2}

\usepackage{graphicx} 
\usepackage{gensymb}
\usepackage{caption}
\usepackage{float}

\usepackage{natbib}
\usepackage{microtype}
\usepackage{amssymb,amsmath}
\usepackage{color}
\usepackage{listings}
\usepackage{wrapfig}
\usepackage{soul}

\usepackage{chemmacros}

\usepackage{upgreek}
\usepackage{enumerate} 
\usepackage[linkcolor = blue, citecolor = blue, urlcolor = blue, colorlinks = true]{hyperref}
\usepackage{hyperref}
\usepackage{xr-hyper} 
\usepackage[english]{babel}
\usepackage{commath}
\usepackage{ushort}
\usepackage{multirow}
\usepackage{cleveref}
\usepackage{epstopdf}

\crefname{equation}{Eq.}{Eqs.}
\crefname{figure}{Fig.}{Figs.}

\usepackage[separate-uncertainty]{siunitx}

\usepackage{hyphenat}
\usepackage{booktabs}

\usepackage{tabularx}
\usepackage{array}
\usepackage{makecell}
\setcitestyle{super,open={},close={}}

\usepackage{xr}
\makeatletter

\newcommand*{\addFileDependency}[1]{
\typeout{(#1)}
\@addtofilelist{#1}
\IfFileExists{#1}{}{\typeout{No file #1.}}
}\makeatother
\usepackage{xr}

\newcommand*{\myexternaldocument}[1]{%
\externaldocument{#1}%
\addFileDependency{#1.tex}%
\addFileDependency{#1.aux}%
}

\myexternaldocument{SI_PNAS}

\newcommand{\affleidenexp}{\affiliation{\small Huygens-Kamerlingh Onnes Laboratory, Leiden University, P.O. Box 9504, 2300 RA Leiden, The Netherlands}}

\mathcode`\,="213B

\begin{document}

\title{{\Large\bf 3D microprinting anisotropic and deformable active matter - A perspective}} 

\author{Mengshi Wei}
\affleidenexp
\author{Daniela J. Kraft}
\email[Corresponding email: ]{kraft@physics.leidenuniv.nl}
\affleidenexp

\maketitle

\section{Abstract}
Active colloidal particles provide versatile model systems for exploring non-equilibrium physics in motile matter. To date, most experimental realizations have focused on spherical particles, largely due to fabrication constraints. However, theoretical and computational studies have long predicted that particle anisotropy and flexibility can dramatically enrich single-particle dynamics, interparticle interactions, and emergent collective behavior.

Here, we highlight recent advances in the fabrication of anisotropic active particles and architectures enabled by the unprecedented design freedom of 3D microprinting. We discuss how additive manufacturing is expanding the accessible parameter space of active soft matter, allowing precise control over shape, location of active forces, and functionality at the microscale. These developments establish new model platforms for uncovering fundamental principles of active and soft matter, and simultaneously pave the way toward microrobotic systems with programmable dynamics and emergent functionalities.
\clearpage

\section{Main text}
\noindent Biological microswimmers, such as bacteria and algae, exhibit a rich spectrum of behaviors, ranging from random exploration to directed navigation in response to environmental cues such as light, chemical gradients, or flow fields. These capabilities enable functions such as foraging, avoidance of harmful conditions, and the formation of complex multicellular structures including colonies and biofilms. At their core, such systems operate far from equilibrium, continuously converting energy into motion and mechanical work. 

Inspired by these living systems, extensive experimental, numerical, and theoretical efforts have been devoted to developing minimal model systems that isolate and elucidate the physical mechanisms underlying motility and collective behavior. Among these, self-propelled colloidal particles have emerged as a particularly powerful platform for studying active matter at the microscale. 
Moreover, their motility promises access to creating microscopic structures capable of emergent collective motion such as swarming, stimulus triggered shape shifting, and even self-healing. 

\noindent\textbf{Colloidal active matter.} Synthetic active colloidal particles combine several key advantages: they are experimentally easily accessible, allow for precise control over interactions and propulsion mechanisms, are small enough to experience thermal fluctuations, and can be tracked at the single-particle level. Most realizations to date have focused on particles with simple, typically spherical geometries, reflecting both fabrication constraints and the desire for minimal model systems. This apparent simplicity has proven remarkably fruitful. Even isotropic self-propelled spheres exhibit a wide range of nontrivial behaviors at both the single-particle and collective level.

Individually, such particles undergo persistent random walks, characterized by diffusive motion at very short time scales, ballistic motion at intermediate time scales, and again diffusive behavior at long time scales.  At the same time, they can interact with each other and their surroundings through a variety of mechanisms, including phoretic, osmotic, steric, electrostatic, and hydrodynamic interactions.\cite{elgetiPhysicsMicroswimmersSingle2015, bechingerActiveParticlesComplex2016} These interactions can be long- or short-ranged, attractive or repulsive, and depend on both the distance and relative orientation with respect to the particle, since the propulsion makes them polar entities. As a result, even spherical active particles can exhibit behaviors such as gravitaxis, alignment in external fields or flows, and complex boundary interactions including wall accumulation and guided motion along interfaces.

At the collective level, these interactions give rise to a wealth of emergent phenomena.\cite{liebchenInteractionsActiveColloids2021} The steric interactions between these spherical polar particles alone can lead to phase separated systems with dense crystalline yet dynamic clusters embedded in a low density of mobile particles.~\cite{palacci2013living, buttinoniDynamicalClusteringPhase2013, theurkauffDynamicClusteringActive2012} - a phenomenon known as motility-induced phases separation which is even robust to the presence of repulsions\cite{cates2015motility}.
In active systems dominated by velocity-alignment interactions, the system undergoes a transition with increasing particle densities from an isotropic gas to collective swarming, flocking, and vortices in confinement.\cite{bricardEmergentVorticesPopulations2015,bricardEmergenceMacroscopicDirected2013}
Together, these findings have established active colloids as a cornerstone model system for non-equilibrium statistical physics.

However, key features of most biological microswimmers but also cells, animals and human beings which are all motile entities, remains largely absent from these minimal realizations: shape anisotropy and deformability or flexibility. In nature, microswimmers are rarely spherical, instead, they exhibit elongated, helical, or otherwise complex geometries that strongly influence their motion, interactions, and functional behavior. Even more so, they all crucially rely on shape changes to break time-reversal symmetry and induce motility at the microscale, for example through the beating of flagella or periodic transformations of their body shape.~\cite{youngSelectiveValueBacterial2006}

From a physics perspective, anisotropy and deformability introduce additional degrees of freedom that can couple translation and rotation, break symmetries, and modify the interactions at the individual and collective level. These features enable direction-dependent forces and torques, non-reciprocal interactions, and feedback between shape, motion, and the environment, all of which can give rise to complex and novel dynamical behavior. Introducing these features in model systems of active colloids will yield key experimental input for developing the physics of thermal out-of-equilibrium systems. 

\begin{figure}
    \centering
    \includegraphics[width=0.5\linewidth]{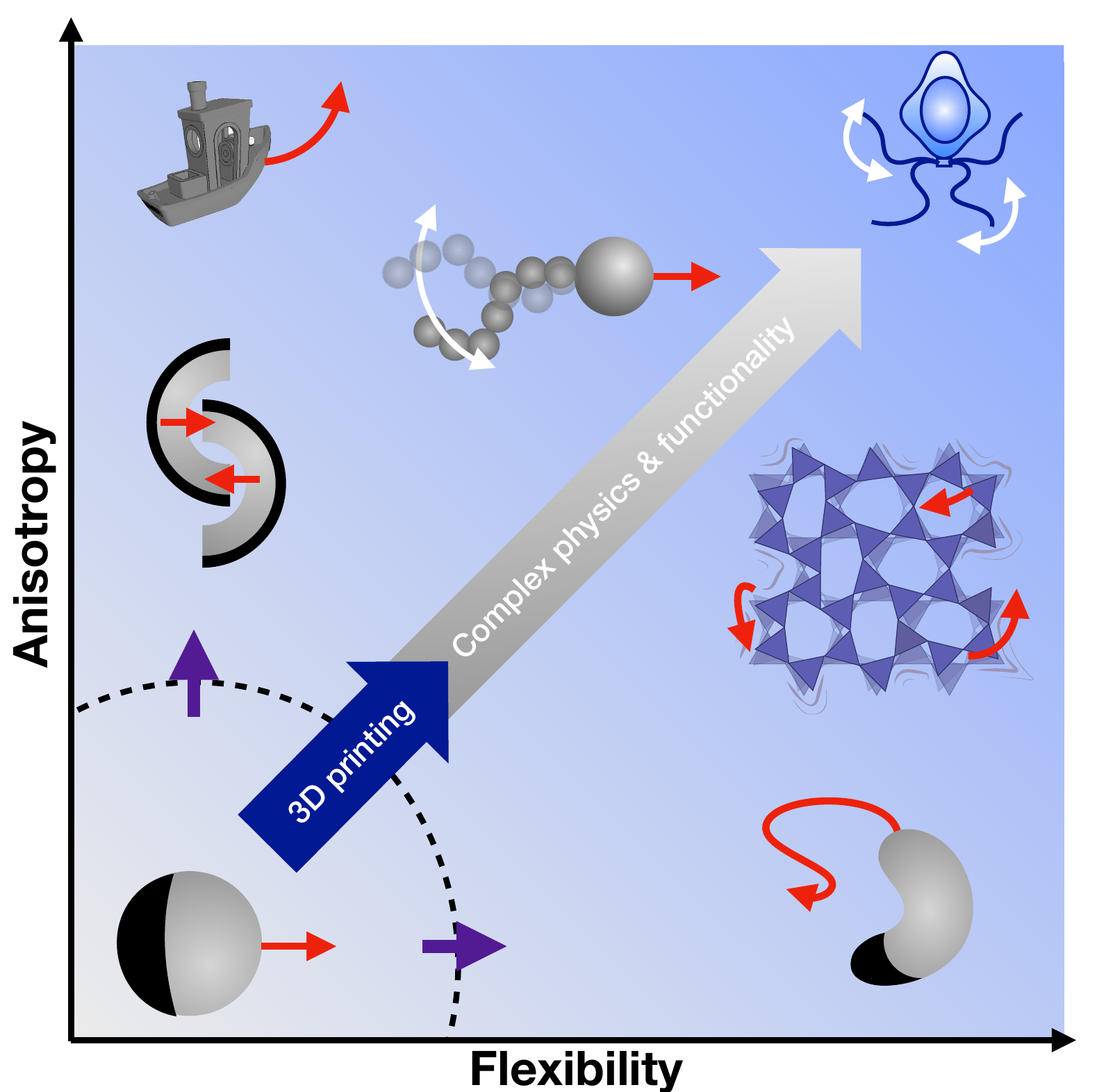}
    \caption{3D microprinting greatly increases the versatility of active matter at the micrometer length scale as it allows integration of shape anisotropy and flexibility. These two design parameters are the key to take research on active colloidal particles from rigid isotropic motile units to complex, intelligent, multi-functional entities and autonomous, adaptive materials with robot-like properties.}
    \label{fig:fig1}
\end{figure}

An understanding of the impact of anisotropy and flexibility in active matter is also important for capturing the behavior of a wide range of biological systems from microswimmers that navigate complex environments to cells and tissues, that actively remodel, sense, and respond to external cues. More broadly, it provides a foundation for designing synthetic active matter in which geometry, mechanics, and dynamics are intrinsically linked, enabling new classes of responsive and functional materials. These materials may possess an ability to adapt their shape to environmental input, process and store information in their shape, and use this information as well as past experiences for taking decisions,\cite{sittiPhysicalIntelligenceNew2021} thereby opening the door towards developing functional materials and devices as well as autonomous, micrometer-sized robots (Fig.\ref{fig:fig1}).

\noindent\textbf{Anisotropic and deformable active matter at the microscale.}
Anisotropic particle shapes have been predicted to significantly enrich the behavior of active systems, mirroring their role in passive colloidal matter.\cite{glotzerAnisotropyBuildingBlocks2007} At the level of individual particles, this additional symmetry breaking of the shape gives rise to a broader spectrum of motion beyond persistent straight trajectories, including circular, helical, and spiral paths, as well as more complex behaviors such as transient dynamics, periodic orbits, and switching between distinct motility modes.\cite{kummel2013circular, wittkowskiSelfpropelledBrownianSpinning2012, shelke2019transition, shemiSelfPropulsionActiveMotion2018a, wang2019active} The direct link between the shape and the type of motion implies that flexibility of the shape can enable different modes and complex dynamics such as self-oscillations and run-and-tumble motion~\cite{ohtaDynamicsDeformableActive2017, winklerPhysicsActivePolymers2020} Coupling between shape, motion, and elasticity has been furthermore predicted to lead to elasto-active instabilities, such as oscillations, symmetry breaking and twisting, as well as quasi-periodic and chaotic motions, and, of course, self-propulsion itself.\cite{ohtaDynamicsDeformableActive2017} 
A more complex, extended shape also allows probing spatial variations more effectively, thereby enabling or enhancing a functional responses to external stimuli and gradients as well as interactions with confining walls.\cite{tenhagenGravitaxisAsymmetricSelfpropelled2014, liebchenViscotaxisMicroswimmerNavigation2018} 

At the collective level, the combination of activity and anisotropic shape gives rise to directional steric interactions, including torques and alignment, which lead to a large variety of non-equilibrium phenomena. 
These have been predicted to include swarming, aligned fronts and rotating clusters, alteration of their phase behavior, including shifting or suppression of MIPS and other forms of dynamic clustering, topological defects, and the emergence of active turbulence and spatio-temporal chaos.\cite{wensinkEmergentStatesDense2012, wensinkDifferentlyShapedHard2013,cugliandoloPhaseCoexistenceTwoDimensional2017, vandammeInterparticleTorquesSuppress2019, barSelfPropelledRodsInsights2020,  rebochoEffectAnisotropyFormation2022, moranParticleAnisotropyTunes2022}. Already the simplest anisotropic shape, an active rod, shows this wealth of nonequilibrium phenomena.~\cite{wensinkEmergentStatesDense2012, barSelfPropelledRodsInsights2020}
Additional complexity arises in collectives of active flexible particles: shape changes upon collision can make alignment easier and accelerate long-range orientational ordering as well as lead to the formation of traveling bands, soliton-like behavior, and active jamming.\cite{menzelSoftDeformableSelfpropelled2012,henkesActiveJammingSelfpropelled2011,ohtaDynamicsDeformableActive2017, manningEssayCollectionsDeformable2023a}. 
 
Despite many exciting theoretical and numerical predictions, experimental exploration of anisotropic active particles remains comparatively limited.\cite{wangEngineeringShapesActive2022} Existing studies have demonstrated complex single-particle dynamics from circular motion to rotations and periodic trajectories, and motion with transitions between distinct motility states \cite{kummel2013circular,  ma2015electric, shelke2019transition, wang2019active, brooks2019shape, caipa2025fabrication}. Side-propelling rods and disks exhibited the formation of hydrodynamically-stabilized pairs of particles, dynamic clusters and low-density phases. \cite{vutukuri2016dynamic,katuri2022arrested} 
Colloidal rods activated by Quincke rotation can make use of a density-dependent feedback mechanism analogous to quorum sensing to switch between a motile and an immotile state, giving rise to absorbing phase transitions and dynamically arrested states.\cite{lefrancQuorumSensingAbsorbing2025}
However, a broad, systematic exploration of the vast parameter space enabled by anisotropic shape is still lacking. 

The reason for the difference between the wealth of theoretical predictions and scarcity of experimental realization lies in the previously limited availability of easily adjustable fabrication methods for active anisotropic particles. 
Synthetic anisotropic active colloids are typically produced via a two-step process: first, passive anisotropic colloidal particles are fabricated and subsequently turned active, which possibly requires further modification depending on the propulsion method. Both steps impose constraints on the achievable active particle characteristics. 

Chemical synthesis allows for easy, scalable fabrication of spherical particles and a limited set of anisotropic shapes including rods, dumbbells, cubes, and patchy particles. More complex shapes such as clusters of spheres or ellipsoids have been obtained through post-synthesis assembly or particle modification, but the structural diversity remains limited. Lithography allows fabrication of particles with more complex shapes, but is largely restricted to quasi two-dimensional geometries.

Incorporating flexibility or deformability in synthetic active matter at the particle level has proven an even more challenging aim. So far, very few experimental realizations exist that combine activity and flexibility. These  include active particles that consist of two materials, of which adapts its size to environmental changes~\cite{alvarezReconfigurableArtificialMicroswimmers2021}, and biological systems such as actin filaments.~\cite{sciortinoActiveMembraneDeformations2025} Another approach to achieve active deformable structures is to connect active particles, for example using light, activity-induced interactions, or enclose them in lipid vesicles.\cite{vutukuriActiveParticlesInduce2020, schonhoferCollectiveBehaviorFlexicles2025, martinetEmergentDynamicsActive2025b, weiReconfigurationInterruptedAging2023} These experiments provided a glimpse of the exciting complex dynamic behaviors that can arise in active flexible systems. However, active particles with an easily deformable shape which adjusts to the environment or changes upon interactions with other particles is essentially missing, limiting the systematic exploration of this deformability in experimental active matter model systems. 

In addition to limitations in the particle shape and deformability, controlling the direction of the active force with respect to the anisotropic particle shape remains a challenge as well. For example, activity can be introduced by an active catalytic patch applied by 2D deposition techniques such as sputter coating or vapor deposition. This inherently limits placement of the active region and furthermore can results in non-uniform shadowing effects from nearby deposited particles. As a result, independently tuning the particle shape and the propulsion direction, which is a key requirement for controlling the behavior of the active particles, is often not possible.

\begin{figure}
    \centering
    \includegraphics[width=0.8\linewidth]{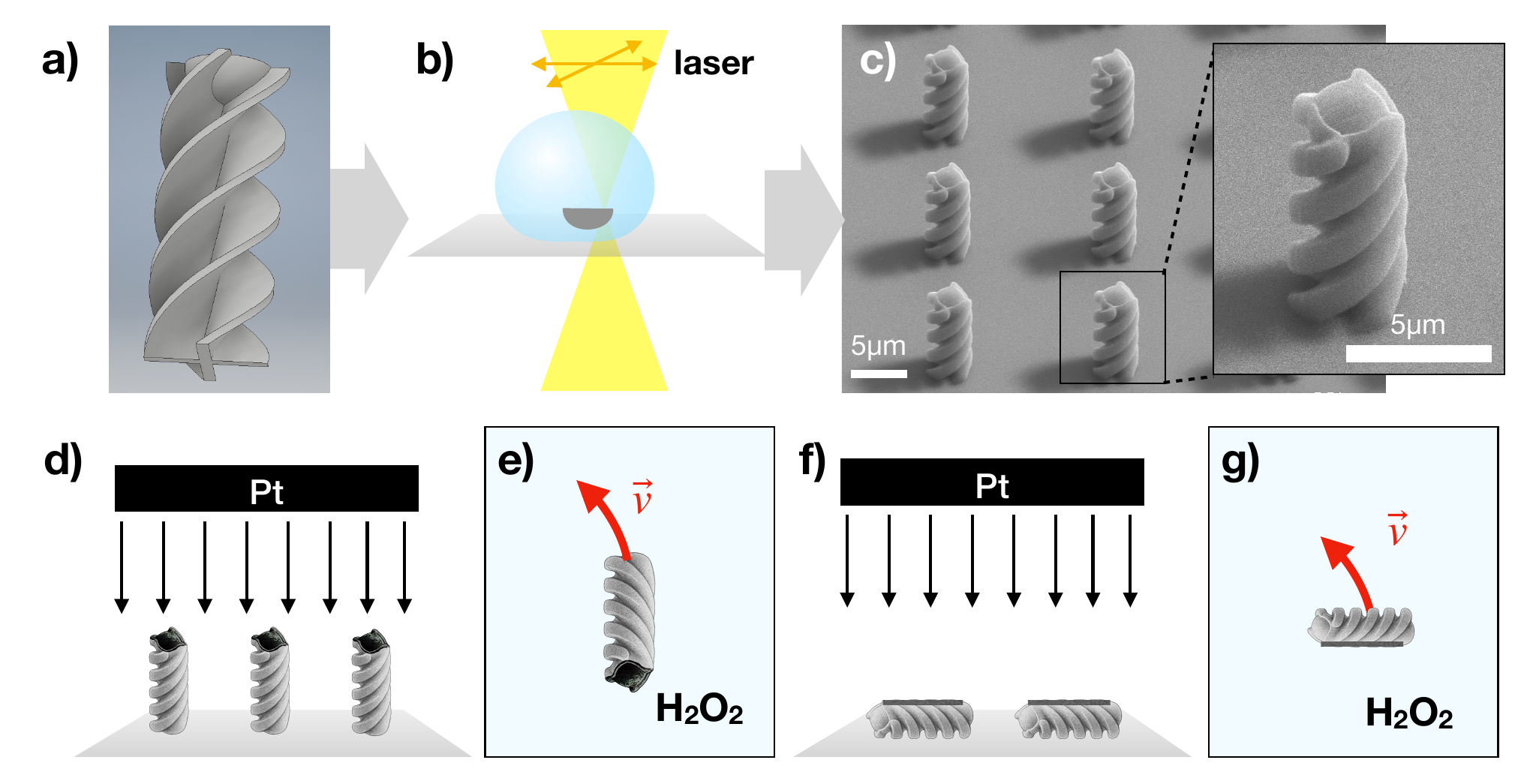}
    \caption{\textbf{3D microprinting}: a) The CAD design is used to guide b)  a focused laser inside a photosensitive material which writes the structure by two-photon polymerization with c) high resolution. SEM image of the written structures. d-g) One approach to render the printed particles active is to apply a catalytic patch by sputter coating, which e, g) leads to propulsion of the particle. 3D microprinting provides a means to control the orientation of the printed particles with respect to the substrate, thereby allowing control over the propulsion direction of the active particle.}
    \label{fig:fig2}
\end{figure}

\noindent\textbf{3D Microprinting offers great versatility}
Additive manufacturing offers an alternative route for fabricating complex microscale structures with unprecedented design freedom. In particular, direct laser writing (DLW) via Two-Photon Polymerization (2PP) has emerged as a the most versatile technique for creating colloidal particles with complex 3D shapes. By tightly focusing a laser into a photosensitive material, polymerization can be triggered locally through nonlinear absorption, enabling voxel-by-voxel fabrication with lateral feature sizes down to 100–200 nm and vertical resolution on the order of a few hundred nanometers. See Figure \ref{fig:fig2}. This allows the realization of micrometer-scale particles with sub-micrometer-sized features whose complexity is limited primarily by connectivity constraints of the printed voxels (Fig.\ref{fig:fig2}c). Both the printed voxels need to be connected to each other and the overall structure needs to be well-attached to a substrate to avoid loss of unattached parts in solution. Although inherently serial and therefore slower than bulk synthesis methods, 2PP offers a high level of structural control that is unmatched by conventional techniques.

\begin{figure}
    \centering
    \includegraphics[width=1\linewidth]{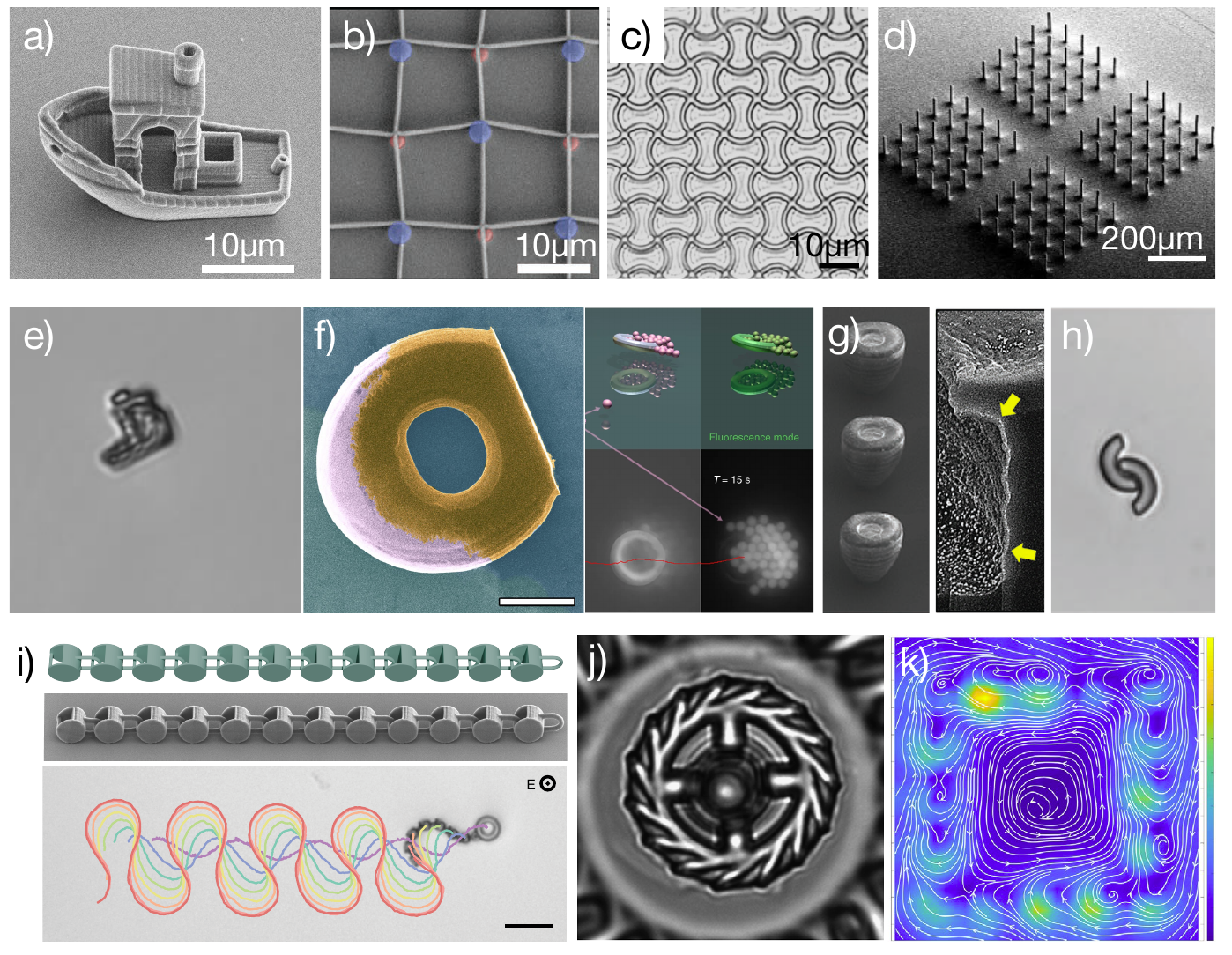}
    \caption{First row: 3D microprinting of a) colloids with a complex shape, reproduced from Ref.~\citenum{dohertyCatalyticallyPropelled3D2020}, b) lattices with integrated colloids, reproduced from Ref.~\citenum{vankesterenPrintingParticlesCombining2023}, c) metamaterials, reproduced from Ref.~\citenum{zhangHydrogelMusclesPowering2023a}, licensed under CC BY 4.0., d) cilia arrays, reproduced from Ref.~\citenum{liu3DprintedLowvoltagedrivenCiliary2026}, licensed under CC BY 4.0. Second row: Examples of active 3D printed particles: e) catalytically self-propelled boat (corresponding SEM image in panel a), and f) torus, reproduced from Ref.~\citenum{bakerShapeprogrammed3DPrinted2019}, licensed under CC BY 4.0, g) nanoparticle modified hydrogel microswimmer, reproduced from Ref.~\citenum{ceylan3DChemicalPatterning2017} Copyright © Wiley and h) bent rods that assembled into a pair, reproduced from Ref.~\citenum{riedelDesigningHighlyEfficient2024}, licensed under CC BY 4.0. Besides propulsion, these anisotropic particles have been shown to be able to f) collect passive particles\citenum{bakerShapeprogrammed3DPrinted2019}, and h) possess different clustering dynamics than active spheres, reproduced from Ref.~\citenum{riedelDesigningHighlyEfficient2024}, licensed under CC BY 4.0. Third row: Integrating flexibility allows i) the development of autonomous microrobots with complex motion patterns and sense response abilities, reproduced with permission from Ref.~\citenum{wei2026life}, j) the realization of microscopic gears where rotation is driven by E. coli, reproduced from Ref.~\citenum{vizsnyiczaiLightControlled3D2017}, licensed under CC BY 4.0, and k) pumping of fluids through artificial cilia, reproduced from Ref.~\citenum{liu3DprintedLowvoltagedrivenCiliary2026}, licensed under CC BY 4.0.}
    \label{fig:fig3}
\end{figure}

\noindent\textbf{Passive anisotropic structures: charting the large unexplored design space}
Since its introduction, the technique has been successfully applied to the fabrication of passive colloidal particles \cite{saraswatShapeAnisotropicColloidal2020,  mayaraniLifetimeFluctuationsSpecific2025} and structures with complex shapes \cite{di2010bacterial}, clusters, crystals and lattices\cite{liu3DPrintingColloidal2022,vankesterenPrintingParticlesCombining2023, mayaraniLifetimeFluctuationsSpecific2025} (Fig. \ref{fig:fig3}). It has furthermore been used to create custom-tailored environments such as obstacle arrays and confining geometries\cite{reinken2020organizing,ketzetzi2022activity}.

Despite these first demonstrations, the potential of 3D microprinting for passive colloidal systems has only begun to be exploited. Self-assembly and phase behavior is highly sensitive to the particle shape, yet experimental studies have so far been limited to a small set of geometries accessible by chemical synthesis.~ \cite{vandenpolExperimentalRealizationBiaxial2009,rossiCubicCrystalsCubic2011,fernandez-ricoShapingColloidalBananas2020} 

The advent of 3D microprinted anisotropic colloids now opens the door to systematically exploring how shape governs interactions, self-assembly, kinetic pathways, phase behavior and transitions, rheological and mechanical properties, and experimentally addressing many more fundamental soft and hard condensed matter questions. Particularly promising are low-symmetry particles not accessible by synthetic methods and families of shapes that can be tuned continuously such as the transition from spheres to ellipsoids and rods. The availability of such particles enable controlled studies of how, for example, symmetry or aspect ratio influence the emergent order, properties, and behavior. Moreover, particle shape can be used to locally encode attractive or repulsive interactions, for example through shape-dependent entropic interactions \cite{damascenoPredictiveSelfAssemblyPolyhedra2012} or by tuning the shape and roughness to create patchy interactions in depletion mediated assembly.\cite{sacannaLockKeyColloids2010, kraftSurfaceRoughnessDirected2012, liuTunableAssemblyHybrid2020,mayaraniLifetimeFluctuationsSpecific2025} 
In addition, post-printing modifications could be utilized to create repulsive or attractive patches following strategies previously developed for synthetically fabricated colloids. 

A key challenge for many studies requiring high densities or large numbers of particles remains the limited throughput of direct laser writing due to its inherently serial nature. However, emerging parallelization strategies, such as multi-focus writing, offer a path toward scaling fabrication, with throughput increasing approximately linearly with the number of writing beams\cite{gu3DNanolithographyMetalens2025,rietz2025dynamic}.

\noindent\textbf{Flexibility and architected materials}
Beyond rigid building blocks, 3D microprinting offers the possibility to create particles with internal degrees of freedom and programmable mechanical compliance. Such deformable units could respond dynamically to interactions with each other or their environment, exciting for understanding soft biological systems where shape changes or deformability are integral for their function, including molecules such as enzymes and collectives of soft entities such as ciliary arrays and tissues.  \cite{henkesActiveJammingSelfpropelled2011,ohtaDynamicsDeformableActive2017,leroyCollectiveDeformationModes2025, liu3DprintedLowvoltagedrivenCiliary2026} 

3D microprinting not only enables precise control over the geometry of individual particles, but also provides a powerful route to create larger structures, where the individual particles are directly assembled into functional architectures with internal degrees of freedom. Functionality may be obtained by combining materials with different mechanical or thermal properties\cite{vankesterenPrintingParticlesCombining2023} or by printing  flexible structures `in situ'.\cite{zhou3DPolycatenatedArchitected2025, wei2026life, liu3DprintedLowvoltagedrivenCiliary2026}
Such flexibility can be realized by employing soft, for example hydrogel-based, materials or through the direct fabrication of mechanical linkages, such as rotational hinges, torsional joints, and multi-axis pivots, that connect otherwise rigid units. The mechanical properties and design of the connections defines the degrees of freedom between neighboring units, enabling controlled bending, twisting, and rotation. This will allow realization of microscale structures with properties akin to macroscopic mechanical meta-materials and origami, with soft deformation modes and functions that are encoded directly into their architecture.\cite{bertoldiFlexibleMechanicalMetamaterials2017, kadic3DMetamaterials2019,  zhangLiquidCrystalElastomerActuatedReconfigurableMicroscale2021,zhangHydrogelMusclesPowering2023a,melioPivotingColloidalAssemblies2026}  The resulting structures are exciting for developing smart materials and integration in functional devices and microrobots, as well as application as powerful model systems for flexible biological structures enabling exploration of the influence of thermal motion on their properties.

\subsection{Active 3D printed particles}
More recently, the capabilities of 3D microprinting have begun to be used in active matter systems. 3D microprinting enables the straightforward fabrication of highly anisotropic particles with a programmable, shape-dependent propulsion mode. Even more so, it also allows orientation of the particles with respect to the substrate during printing, advantageous for propulsion mechanisms that rely on the precise application of a patch on the particle surface with 2D deposition techniques. Controlling the orientation of the particles, hence controls the location of the patch and thus the direction of the propulsion force with respect to the particle (Fig.\ref{fig:fig2}d-g).\cite{bakerShapeprogrammed3DPrinted2019, dohertyCatalyticallyPropelled3D2020} Selective 3D modification of the particle surface by two-photon crosslinking after printing even allows adhesion of (catalytic) nanoparticles to any desired site (Fig.\ref{fig:fig3}g).\cite{ceylan3DChemicalPatterning2017} 

The versatility of 3D microprinting provides access to systematically investigating how shape affects the motion, interactions, and collective behavior of active matter. Since additive manufacturing provides independent control over particle geometry and placement of the propulsion direction, it allows design of active particles with the propulsion axis  precisely oriented with respect to the anisotropic body. Because particle geometry directly determines the hydrodynamic friction tensor, this provides a powerful means to program motion at the single-particle level and test theoretical predictions.\cite{wittkowskiSelfpropelledBrownianSpinning2012}
Initial experiments already illustrate the potential: helical particles, for example, exhibit a coupling between translation and rotation, resulting in screw-like propulsion.\cite{dohertyCatalyticallyPropelled3D2020} Anisotropic catalytically-active particles such as crescents, disks and tori can exhibit direction reversal of their motion that depends on the type of catalytic coating and fuel concentration~\cite{bakerShapeprogrammed3DPrinted2019,riedelShapedependentDirectionReversal2025} - where an anisotropic shape affects the sensitivity to environmental conditions by influencing the confinement of solutes.~\cite{michelinGeometricTuningSelfpropulsion2017a} 3D printing active particles offers the opportunity to systematically explore shape-dependent motility, where in particular families of shapes with continuously tunable geometry might provide valuable general insights\cite{riedelDesigningHighlyEfficient2024} and low-symmetry particles might exhibit the most complex motion patterns.  

Beyond catalytic swimmers, 3D microprinting also provides a platform to integrate a broad spectrum of propulsion mechanisms, including induced charge electrophoresis, dielectrophoresis, Quincke rotations, demixing solvents, or thermophoresis. Recent work demonstrates that even anisotropic shapes alone can induce propulsion, which makes direct programming of the propulsion force and strength through the shape possible.\cite{brooks2019shape, wei2026life}
Further modification with magnetic particles or layers can provide external control over the motion and orientation of the particles.\cite{bakerShapeprogrammed3DPrinted2019, smartMagneticallyProgrammedDiffractive2024} Finally, 3D microprinting can also bridge the gap from microscopic to inertial active system, by allowing the fabrication of particles of similar shapes yet different sizes.\cite{lowenInertialEffectsSelfpropelled2020}

The great design freedom also  opens the door to studying how anisotropic active particles behave in complex environments and confinement, respond to external fields and gradients. While various predictions have been made as highlighted above, experiments are largely missing, and the potential for discovering unexpected behaviors in these out-of-equilibrium systems is great. For example, an anisotropic particle shape such as a cube can have different stable orientations with respect to a substrate due to interaction with the gravitational field, leading to distinctly different motilities within an otherwise uniform population of active particles.\cite{caipa2025fabrication}

Moreover, the availability of designable active particles will allow addressing fundamental question as to how shape affects their active phase behavior and phase transitions, and using these results to validate, refine, and develop theoretical models as well as to unravel the effect of motility by comparing them to the analogous passive systems. For example, the formation of clusters already appears at much lower densities for active spheres than for passive spheres.\cite{palacci2013living, buttinoniDynamicalClusteringPhase2013, theurkauffDynamicClusteringActive2012} However, active particles with more complex, interlocking geometries such as bent rods can cluster with high efficiency at arbitrarily low densities, and their shape furthermore controls the timescale of the assembly kinetics.~\cite{riedelDesigningHighlyEfficient2024}. Tuning the shape thus can provide another means to control the interactions and assembly pathways of active particles, with great potential in bottom-up self-assembly of functional materials. Exploring the collective out-of-equilibrium behavior of active particles thus will likely lead to rich new physics and dynamic phenomena. 

Finally, tuning the shape of active and passive micrometer-scale units can also be exploited for the integration of functions. Besides complex motion in 2D and 3D, these may include navigation of complex environments, transport of other objects and particles ~\cite{bakerShapeprogrammed3DPrinted2019}, and dynamic pumping of fluids such as was recently demonstrated with ciliary arrays.\cite{liu3DprintedLowvoltagedrivenCiliary2026} Shape-directed interactions can even be exploited to guide the assembly of active and passive particles into functional micromachines,~\cite{alapanShapeencodedDynamicAssembly2019} where the activity can provide the energy to power functional shape changes and thus do work.

\subsection{Active solids and metamaterials at the microscale}
Active solids (or active metamaterials) composed of active units connected by soft interactions have been shown to possess rich emergent behaviour at the macroscale~\cite{baconnierSelectiveCollectiveActuation2022}. In these systems, individual active units are elastically coupled so that mechanical stresses propagate through the network, allowing local activity to organize into collective modes such as coherent motion, synchronized oscillations, or mode-selective actuation\cite{ferrante2013elasticity,baconnierSelectiveCollectiveActuation2022,veenstra2025adaptive}. 3D microprinting offers a powerful route to extend this concept to the microscale and fabricate and program active solids and active metamaterials with thermal fluctuations. It provides precise control over anisotropic geometry and connectivity, enabling activity to be encoded directly into the architecture.

Interestingly, translating this concept to the microscale introduces additional physical interactions that might fundamentally reshape the observed behaviors. At micron scales, active units interact not only through their mechanical links or flexible bonds but also through the surrounding fluid, charges, and chemical fields. 
Hydrodynamic flows decay slowly in low-Reynolds-number environments, mediating long-range coupling across the system which can drive collective phenomena observed in active suspensions, including coherent flocking states, vortex-like swirling flows, hydrodynamic instabilities, and large-scale pattern formation\cite{bricard2013emergence,dunkel2013fluid, saintillan2008instabilities}.
Chemical interactions introduce another layer of coupling between active units. Active particles interact through a shared reaction–diffusion field: each particle both generates chemical gradients and responds to those produced by others, creating a feedback loop that mediates long-range interactions and collective dynamics. Such coupling can lead to behaviors observed in active colloidal systems, including dynamic clustering (living crystals), chemotactic aggregation and non-equilibrium phase separation\cite{palacci2013living, stark2018artificial, agudo2019active}.

Consequently, an elastically connected architecture composed of active building blocks effectively becomes an elastic structure embedded within a self-generated hydrodynamic, electrostatic and chemical field. These interactions are often non-reciprocal and can break conventional action–reaction symmetry, providing a microscopic route toward unusual material responses such as odd elasticity.\cite{scheibnerOddElasticity2020}

Moreover, anisotropic geometries of the constituent units, readily achievable through 3D microprinting, allow active forces to generate not only translation but also torques, bending moments, and twisting stresses within the elastic network. A similar mechanism operates in biological flagella, where molecular motors do not simply pull the filament forward but instead generate distributed bending moments along the filament, producing self-organized traveling waves.

The interplay between elasticity, anisotropic activity, fluid flows, and chemical gradients may therefore give rise to feedback mechanisms and collective behaviours far beyond those observed in macroscopic robotic active solids, such as spontaneous oscillations, traveling deformation waves, adaptive, continuous or hysteretic shape morphing.

\subsection{Expanding the materials for 3D printing}
Beyond conventional polymer resins, the range of materials available for 3D microprinting is rapidly expanding to resins with diverse physical and chemical functionalities. Emerging photoresists based on hydrogels, preceramic precursors, conductive nanocomposites, and stimuli-responsive materials already demonstrate that microprinted architectures can incorporate various functions, from mechanical to optical\cite{lyu2026optofluidic,dzikonski2025hybrid,zhou2024photoresist,demirorsThreedimensionalPrintingPhotonic2022a}. 

For the field of active matter, these advances open particularly exciting opportunities. A broader range of printable materials will enable the integration of active elements, responsive components that adapt to their environment, and functional interfaces directly into designed microstructures. This will enable synthetic systems with richer dynamics, programmable responses, diverse sense-reponse functionalities and closer analogies to biological active materials, ultimately advancing the development of more life-like, intelligent microrobotic systems.

\section{Concluding remarks}
3D microprinting is emerging as a powerful technique for fabricating the next generation of active particles and materials with functions embedded through their shape and architecture, placement of active forces, and integration of flexibility and deformability. These active 3D printed structures have great potential as model systems for obtaining quantitative insights into a large range of biological systems, from active polymers, to microorganisms, and tissues. They can furthermore be employed for the developing advanced materials that might be integrated in the next generation of biomedical devices that autonomously adapt to their environmental conditions as well as micrometer-sized robots with autonomous functions, sense-response abilities, material-based memory and information processing. Expanding the range of resins, post-modifications, and speeding up the fabrication through parallelizing the printing will expand the range of applications and impact of this technique even further.

\textbf{Acknowledgments.} We thank Silvana Caipa Cure and Alexandre Morin for constructive feedback on the manuscript. 
\bibliography{reference}

\end{document}